\documentclass[a4paper]{jpconf}
\usepackage{graphicx}
\usepackage{hyperref}

\hypersetup{
    colorlinks=true,       
    linkcolor=red,          
    citecolor=blue,        
    urlcolor=blue           
}

\begin{document}
\title{STAR Data Reconstruction  at NERSC/Cori, an adaptable Docker container approach for HPC}

\author{Mustafa Mustafa$^1$, Jan Balewski$^1$, J\'{e}r\^{o}me Lauret$^2$, Jefferson Porter$^1$, Shane Canon$^1$, Lisa Gerhardt$^1$, Levente Hajdu$^2$,  Mark Lukascsyk$^2$}
\address{$^1$ Lawrence Berkeley National Laboratory, One Cyclotron Rd, Berkeley, CA 94720}
\address{$^2$ Brookhaven National Laboratory, NY 11973}
\ead{mmustafa@lbl.gov}

 \begin{abstract} 
  As HPC facilities grow their resources, adaptation of classic HEP/NP
   workflows becomes a need. Linux containers may very well offer a way to lower
   the bar to exploiting such resources and at the time, help collaboration to
   reach vast elastic resources on such facilities and address their massive
   current and future data processing challenges.  

In this proceeding, we showcase STAR data reconstruction workflow at Cori HPC
  system at NERSC. STAR software is packaged in a Docker image and runs at Cori
  in Shifter containers.  We highlight two of the typical end-to-end
  optimization challenges for such pipelines: 1) data transfer rate which was
  carried over ESnet after optimizing end points and 2) scalable deployment of
  conditions database in an HPC environment. Our tests demonstrate equally
  efficient data processing workflows on Cori/HPC, comparable to standard Linux
  clusters.
\end{abstract}

\section{Introduction} 
The expected growth in High Performance Computers (HPC) capacity over the next
decade makes such resources attractive for meeting future computing needs of
High Energy and Nuclear Physics (HEP/NP) experiments \cite{ascrhep}, especially as their cost is becoming
comparable to traditional clusters.  However, HPC facilities rely on features
like specialized operating systems and hardware to enhance performance that
make them difficult to be used without significant changes to data processing
workflows. Containerized software environment running on HPC systems may very
well be an ideal scalable solution to leverage those resources and a promising
candidate to replace the outgrown traditional solutions employed at different
computing centers.  

While STAR data real-data reconstruction has been carried at remote sites before in a sustainable and geographically distributed manner \cite{kisti}, the HPC environment provided a new set of unique challenges. In this proceeding we report on the first test of STAR real-data reconstruction
utilizing Docker \cite{docker} containers on the data partition of the Cori supercomputer at
NERSC\footnote{\url{http://www.nersc.gov/users/computational-systems/cori/}}. The
target dataset was recorded by the STAR experiment at RHIC in 2016
and is estimated to require $\sim$50~M core-hours for full processing. To ensure
validity and reproducibility, STAR data processing is restricted to a vetted
computing environment defined by system architecture, Linux OS, compiler and
external libraries versions.  Furthermore, each data processing task requires
certain STAR software release and database timestamp. In short, STAR data
processing workflow represents a typical embarrassingly parallel HEP/NP
computing task.  Thus, it is an ideal candidate to test the suitability of
running containerized software, normalized to run on a shared HPC systems
instead of its traditional dedicated off-the-shelf clusters. This direction, if
successful, could address current and future experiments' computing
needs. We report on the different opportunities and challenges of running in
such an environment.  We also present the modifications needed to the workflow
in order to optimize Cori resource utilization and streamline the process in
this and future productions as well as performance metrics.

Section 2 outlines some of the technical aspects of our pipeline and discusses
conditions database location in an HPC environment. In section 3,
we show performance results from an extended test that uses +120k core-hours.
Summary and outlook are provided in section 4. 

\section{Implementation}
\subsection{Docker/Shifter}
HPC systems provide vast resources for computation and data intensive
applications. However, for technical and logistic reasons, their software is
not readily customizable for specific project needs. With a Docker enabled
system, experiments can run their operating system of choice  and a customized
software stack in Linux containers.  The ability to "push" the software image
with the job to the computing nodes makes any computing environment, e.g. HPC
or Cloud, very versatile. Shifter \cite{shifter} is a NERSC project created to
allow easy custom software stack deployment by bringing the functionality of
Linux containers to Cray compute nodes along with the tool to convert Docker
images into Shifter images for running on NERSC HPC systems.

We base our STAR Docker image on Scientific Linux 6.4. ROOT \cite{root} and STAR software
are baked into the image from pre-compiled binaries. MySQL server and OpenMPI
are the only components in the Docker image that are different from the STAR
software environment used on traditional clusters. The MySQL server providing 
conditions DB is launched locally on the compute node. MPI is used to fan-out 
jobs to process sections of the same input data file by as many data
reconstruction chains as available number of processors (or threads) on the
node. This is complete data parallelism without any message passing between threads.

\subsection{Pipeline}

\begin{figure}[h]
\begin{center}
\includegraphics[width=0.9\textwidth]{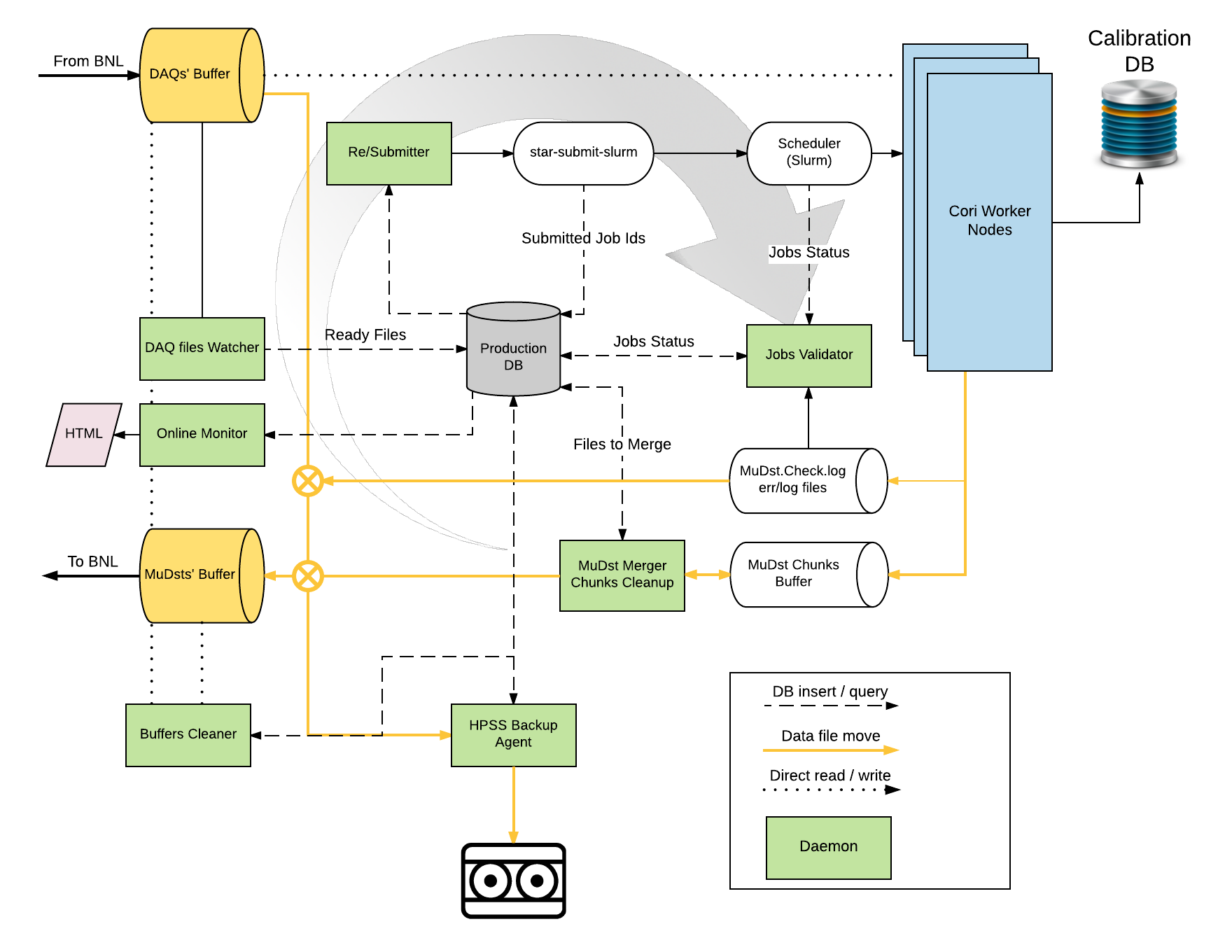}
\end{center}
  \caption{\label{pipeline}A schematic of STAR data reconstruction pipeline at Cori, NERSC. 
  The raw data input files (DAQ's) are transferred from BNL to NERSC over ESnet. The reconstructed data (MuDST's) 
  are pulled back from BNL side.}
\end{figure}

A schematic of the pipeline we built for reconstruction of STAR data at Cori is
shown in Figure \ref{pipeline}. To carry a months long data processing campaign at
an HPC facility requires a high level of pipeline fault-tolerance. In addition,
HPC systems are known for their more frequent maintenance downtimes, thus the
pipeline needs to demonstrate cold-start capability. For these reasons and to
facilitate end-to-end optimization we opted for a modular multi-threaded design
(12 daemon threads) as opposed to a monolithic design. Our pipeline is designed
for a target throughput of 10,000 cpu-cores continuously running to process raw
data in the order of 100~TB per week. The entire campaign is expected to use
25~M core-hours.  The pipeline handles 1) transfer of raw data files (DAQ's)
from BNL to NERSC over ESnet and checking for integrity (md5sum) 2) submitting jobs to
the queue; one file can be processed using 1 to 32 processors depending on
queue partition availability and input file size. The output of the different
processors are later merged into one file (MuDST) and 3) perform quality
assurance on the output of the jobs and transfer successfully produced data back
to BNL. 

Central to our design is the production database to which all the states of the
different stages of input file processing, daemons states (heartbeats) and
summary statistics are logged. The daemons themselves are internally stateless
and no communications between daemons happen except through the database, this
provides 1) persistent storage of all pipeline states 2) daemon failure
tolerance; daemons can be individually restarted and they pickup from where
they left off 3) continuous collection and monitoring of pipeline states and
statistics is easily done by querying the database. We rely on
Slurm\footnote{Slurm Workload Manager is the scheduler used by many supercomputers as well as Cori} \cite{slurm} to
monitor jobs states and resource utilization on the compute node, i.e. no
heartbeats directly from the job to the production DB.

In addition to the core tasks, Figure \ref{pipeline} also shows some auxiliary
daemons, such as HPSS backup agent and the buffers cleaner. A Python Flask app
running on NERSC web-portal responds to web-base users queries to production DB 
and generates aggregate statistics plots showing progress of the production.
This allows the operators of the production campaign to catch failures of the
systems that are not automatically handled by the pipeline daemons.

\subsection{Conditions database}
STAR data processing software uses a MySQL service for detector online running
conditions and calibration parameters. The location of this DB is critical for
scaling the number of jobs to the planned throughput of the pipeline. Given
Cori layout and STAR framework capabilities we had a couple of options for this
DB: 1) a network isolated DB service, i.e. a DB server on a machine outside the
Cori network or 2) full or minimal snapshot payload DB server local to the
compute node.  Other options such as launching DB servers in separate jobs to
serve a collection of compute nodes were entertained but deemed to pose
unnecessary technical challenges. 

We found that network isolated DB service suffers from unpredictable network
conditions, in particular network throughput relies on other users' traffic and
other not easily foreseen conditions. For example, during our stress tests of
the pipeline the network was under performing which made the jobs spend 45\% of
their walltime waiting for the DB. Jobs usually spend less than 2\% of their
walltime in the DB. This incident lasted for more than 24hrs. At this point it
was clear that we have to move the DB server closer to the jobs.

Relying on DB servers local to each computing node comes at a cost.  In
particular, as with many HPC systems, Cori nodes do not have local disks. The
MySQL DB payload has to be placed on a globally mounted Lustre filesystem. 
The Lustre filesystem is not designed for the database transactional workloads that
do a lot of metadata lookups. This is particularly problematic; the need to
start a MySQL server with a fresh cache at the beginning of each job causes many
queries to Lustre metadata to build up the data in-memory. As a result, long delays were
observed during the start of each job due to this inefficient use of the 
filesystem. For example, a particular table requires $\sim$30k queries to
build, building such a table takes more than 30 minutes with the payload on
Lustre. 

Shifter provided a solution to the problem of slow startup times due to
building the DB cache on Lustre with its perNodeCache feature.  This allows
for the creation of a write-able XFS file which is loopback mounted on the compute node
and acts as a local file system for the container.
While the loopback mounted file is placed on Lustre, the metadata can be safely cached in the compute node's memory since it is private to the node.  We opted for a solution where every job starts by copying the MySQL
payload into its own perNodeCache then launches the MySQL server.  To simplify
the workflow we copy the entire STAR DB payload as opposed to a single
year's snapshot, copying the 30 GB payload takes 1-2 minutes. This solution
dramatically changed the DB performance. For example, caching the $\sim$30k
queries table dropped from +30 minutes to a few seconds. We launch one DB
server per compute node which serves all the reconstruction chains on that
node, 1-32 chains on Cori Haswell nodes.  We found that over-committing the node
cores has not decreased the aggregated events throughput per core per walltime.
So we run 1 MySQL server + N reconstruction chains on N cpu-cores. The local DB
server strategy avoids the network routing bottlenecks altogether and trivially
scales to our projected pipeline throughput.

\subsection{NY-CA data transfer}

\begin{figure}[h] 
  \begin{center}
    \includegraphics[width=0.45\textwidth]{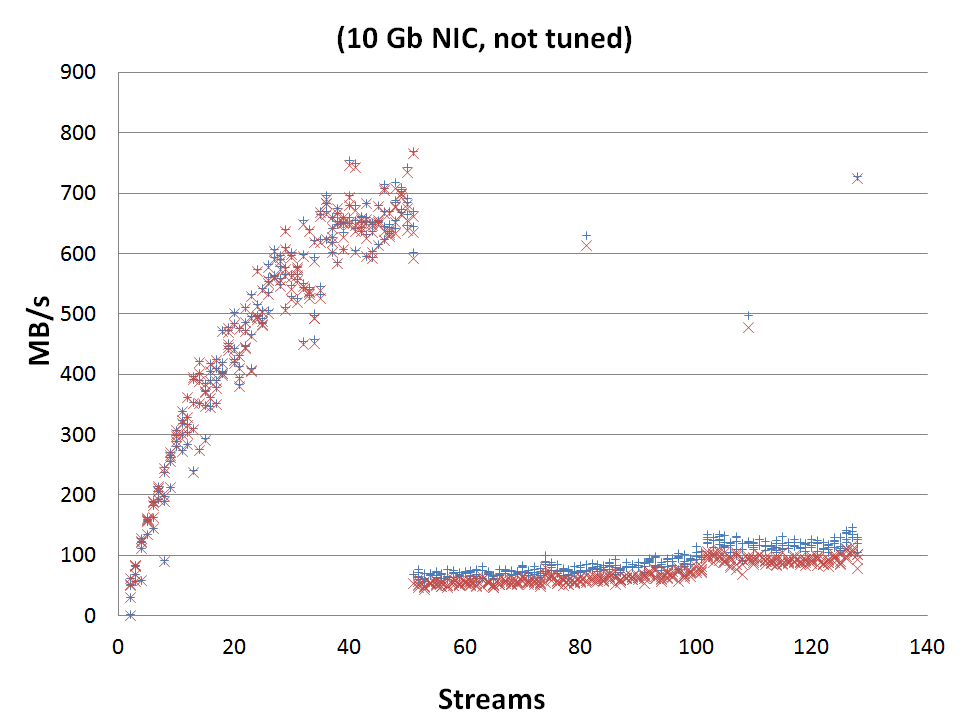}
    \includegraphics[width=0.46\textwidth]{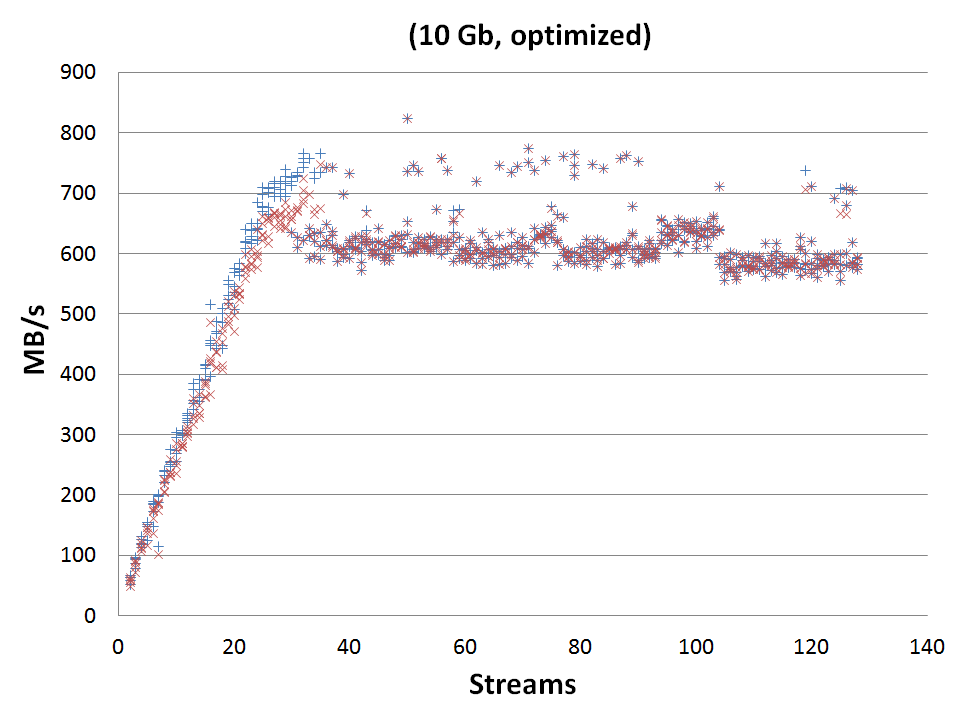} 
    \caption{\label{network}
    Transfer rates over ESnet before (after) end-points and transfer protocols
    optimization, left (right), 5x improvement over the vanilla setup} 
  \end{center} 
\end{figure}

For best utilization of CPU resources we need reliable data transfer between
BNL and NERSC. Projected throughput requires the transfer of $\sim$100 TB/week
and $\sim$15 TB/week of reconstructed data back to BNL or approximately a
continuous transfer rate of 200 MB/s in steady state production. The data is
transferred over the Energy Sciences Network (ESnet)\footnote{\url{https://www.es.net/}}. 

Our initial setup was composed of two active-passive data transfer nodes connected at 2x1 Gb speeds behind a firewall.  This did not fulfill our minimal requirement of 200 MB/sec. We started by upgrading the interface to 10 Gb network interface cards (NIC) and migrated the transfer nodes to BNL's Science DMZ which connects each node at 2x10Gb (LACP). Our initial testing of the performance is showed on Figure \ref{network} (left). The network transfer pattern, showing a bimodal feature, is a clear indication and a sign of transfer collapse  due to congestion and transfer buffer “size”  - following the basic tuning instructions available on \cite{esnet_opt}, we obtained and confirmed stable data transfers with a plateau at 600 MB/sec. The most relevant tuning were the change to the congestion control from bic to htcp and modifying the rmem/wmem parameters as suggested (Fig. \ref{network}, right).

\section{Performance test} 

\begin{figure}[h] \begin{center}
  \includegraphics[width=0.7\textwidth]{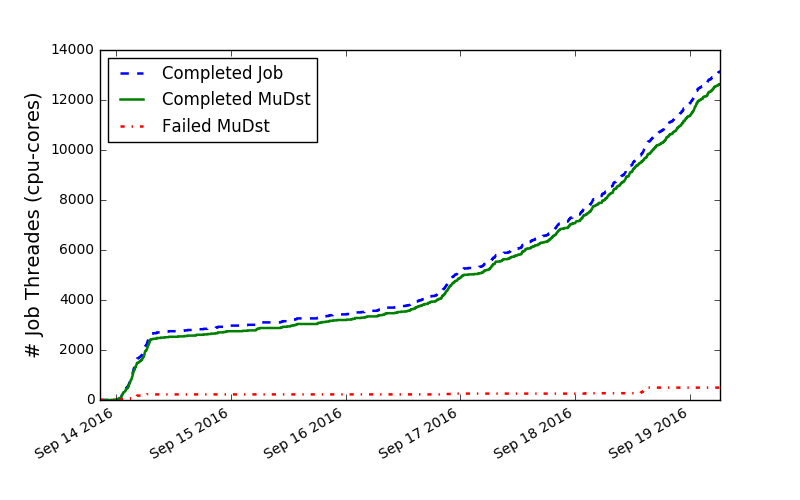}
  \caption{\label{production_mon} An example of pipeline monitor plot showing
  aggregate statistics of successful and completed jobs during a test
  production.  "Completed job" is for jobs which ran to completion. "Completed
  muDst" is for produced files which pass QA and +98\% of events produced}
\end{center} \end{figure}

We carried a real-data reconstruction test that used +120k core-hours to test
the different pipeline units integration and calculate the overall success
rate (SR). Our criteria for a successful output file is for the job to run to 
completion and for the file to have +98\% of the data events
reconstructed.  Figure \ref{production_mon} shows job success/failure aggregate
statistics during our test production which was carried over 6 days. The tests
had a SR $> 96\%$ for reconstructed files that passed our QA. Generally, we
require an SR $> 95\%$ to qualify the workflow and computing facility
to be real-data processing quality. The failed $\sim$4\% are mainly due to the
failure of the local MySQL server. Solving this problem by making the server
persistent will allow us to increase the SR to the +99\% range.

\begin{figure}[h]
  \includegraphics[width=0.46\textwidth]{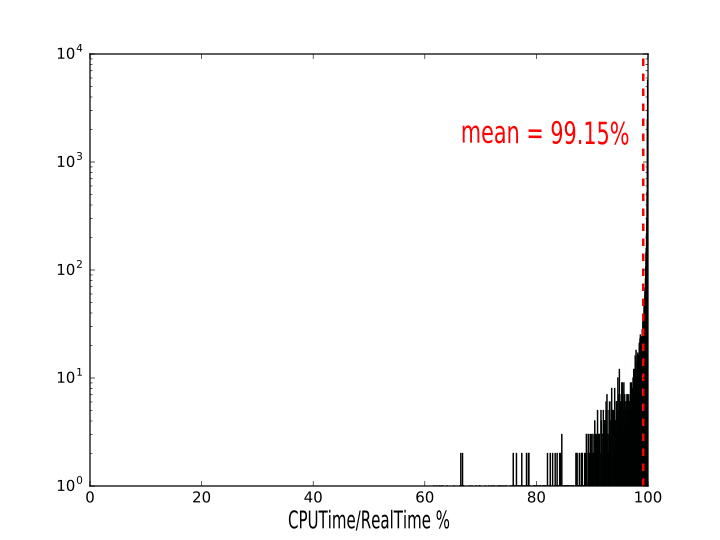}
  \includegraphics[width=0.46\textwidth]{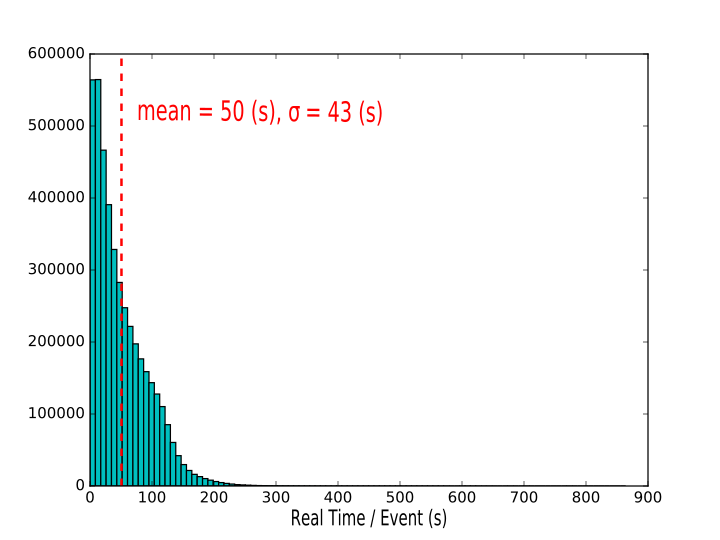}
  \caption{\label{performance} (Left) CPU utilization efficiency
  CPUTime/WallTime (Right) Distribution of walltime to process one event}
\end{figure}

Finally, we examined the CPU utilization efficiency of the jobs.
Each one of our test jobs was granted 16 cpu-cores, and ran 16 reconstruction
chains + 1 MySQL server (overcommitting the cores). The CPU utilization
efficiency is well above 99\% for most jobs as shown on the left plot in Figure
\ref{performance}. Figure \ref{performance} right shows the distribution of
walltime per event. The average time of 50s/Au+Au event is comparable to 48s
without a local MySQL server. 

\section{Summary \& Outlook}
In this proceeding we outlined STAR strategy for carrying HEP/NP real-data
reconstruction on Cori HPC facility. Docker/Shifter-enabled Cori and Edison at
NERSC have vast resources that can be used to run customized HEP/NP software
stacks. Edge services availability, such as conditions database, are a
traditional challenge for running in HPC environments. We showed that, in the
database case, running a database server at the compute node with the DB
payload in a Shifter perNodeCache is a highly scalable solution without much
overhead on the CPU resources. ESnet enables the transfer of large amounts of
data across the continent; however, we found that both network protocols and
endpoint optimizations are essential components of steady-state end-to-end
transfer operations.  Finally, our test production shows +99\% walltime with
very high success rate (+95\%), allowing us to deem the pipeline at Cori
real-data processing quality. We plan to use the pipeline to process half of
the 3Pb of Au+Au collision data recorded by the STAR experiment during RHIC run
2016 which  will require $\sim$25 M core-hours. 

\section{Acknowledgements}
We thank the RHIC Operations Group and RCF at BNL, the NERSC Center at LBNL, and the Open Science Grid consortium for providing resources and support. This work was supported in part by the Office of Nuclear Physics within the U.S. DOE Office of Science.

\end{document}